# 4D-STEM elastic stress state characterisation of a TWIP steel nanotwin

**T P McAuliffe*[a]**, A K Ackerman[a], B H Savitzky[b], T W J Kwok[a], M Danaie[c], C Ophus[b], D Dye[a]

*t.mcauliffe17@imperial.ac.uk

[a] Department of Materials, Prince Consort Road, Imperial College London, UK
[b] National Center for Electron Microscopy (NCEM), Molecular Foundry, Lawrence Berkeley Lab, USA
[c] Electron Physical Science Imaging Centre (ePSIC), Diamond Light Source, UK

**Abstract:**

We measure the stress state in and around a deformation nanotwin in a twinning-induced plasticity (TWIP) steel. Using four-dimensional scanning transmission electron microscopy (4D-STEM), we measure the elastic strain field in a 68.2-by-83.1 nm area of interest with a scan step of 0.36 nm and a diffraction limit resolution of 0.73 nm. The stress field in and surrounding the twin matches the form expected from analytical theory and is on the order of 15 GPa, close to the theoretical strength of the material. We infer that the measured back-stress limits twin thickening, providing a rationale for why TWIP steel twins remain thin during deformation, continuously dividing grains to give substantial work hardening. Our results support modern mechanistic understanding of the influence of twinning on crack propagation and embrittlement in TWIP steels.

Understanding the mechanical behaviour of engineering alloys at the nanoscale is critical to improving alloy design and processing, and hence performance [1–4]. Plastic deformation by twinning or shear-associated martensite transformation is commonly used in alloy design strategies, from Mg to Zr, steels and body-centred cubic (BCC) Ti alloys, as well as functional intermetallics such as NiTi. A better understanding of the stress state and back stress around such features will open up new avenues for improving performance, by manipulation of composition and processing to achieve desirable nanoscale behaviour and therefore bulk properties. In the case of face-centred cubic (FCC) crystal twinning, Shockley partial <112>{111} dislocations formed from the dissociation of a full <110>{111} lattice dislocation propagate on successive {111} planes. These impart a local strain field, and their propagation (and therefore accumulation of plastic strain) is limited by interaction with other lattice defects (interstitial or substitutional solutes, additional twins, dislocations, *etc*).

Four-dimensional scanning transmission electron microscopy (4D-STEM) is a relatively new technique in which an electron diffraction pattern is acquired at every point in a scan grid. In this regard it is similar to electron backscatter diffraction (EBSD), now a routine method for microscale structural analysis [5,6], but a much finer 'pencil-beam' probe permits sub-nm spatial resolution. The trade-off is that a zone-axis generally must be identified and aligned with the transmitted beam, inherently limiting knowledge of the reciprocal lattice to two coplanar vectors. A comprehensive review of 4D-STEM and its applications in strain mapping, imaging, and ptychography is available in ref [7]. In this work we employ the py4DSTEM open source software package, developed by Savitzky *et al* [8].

Lattice strain measurement with this approach is becoming fairly routine. It has been used to investigate resistivity in semiconductors [9,10], and more recently begun to be applied to polycrystalline materials [11]. Pekin *et al* [12,13] have measured the strain field around austenitic (FCC) stainless steel features. They observed a ~4% variation in strain across their area of interest, which included dislocations and an annealing twin boundary. In this study we investigate the elastic strain fields in a similar FCC Fe material, and additionally calculate the stress fields directly from the elastic strain measurements. Here we examine twinning-induced plasticity (TWIP) steel, with a focus on thin deformation nanotwins. Twinning-induced plasticity in these systems can result in large ductility of up to 95% [14–16]. It is believed that continuous subdivision of grains by ongoing twin nucleation, without significant thickening, leads to a dynamic 'Hall-Petch' effect with sufficient twin back-stress to inhibit propagation of dislocations at these barriers [17]. This leads to pile-ups, hardening, and ductility. Deformation twins have also been explored in the context of crack initiation in TWIP steels by Koyama *et al* [18]. In this study we provide a direct measurement of the nanoscale stress state for comparison to the increasingly cited analytical model of Müllner *et al* [19–21]. We observe that the analytical form of this model qualitatively corresponds well to our measurements of local stress.

An ingot of TWIP-steel (Fe – 16.4Mn – 0.9C – 0.5Si – 0.05Nb – 0.05 V wt%) was produced by vacuum arc melting in an Ar atmosphere. It was cast, homogenised at 1300°C for 24 h, hot rolled with a 50% reduction, cold rolled with a subsequent 67% reduction, and annealed at 1000°C for 5 min. This gave a fully austenitic microstructure. A 'dogbone' tensile specimen with 1 x 1.5 mm cross section and 19 mm gauge was deformed to 6% plastic strain at a nominal stroke rate of $10^{-3}$ s$^{-1}$. The material in this condition has a yield strength of 1.07 GPa, ultimate tensile strength of 1.5 GPa, ductility of 33%, and work hardening rate of 3.5 GPa.

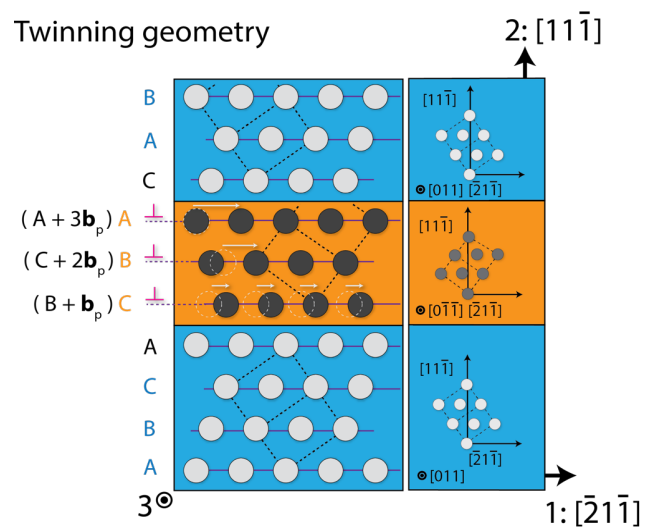

**Figure 1**: Twinning in FCC materials, with habit plane (11$\bar{1}$), and [011] zone axis. A series of Shockley partial [$\bar{2}$1$\bar{1}$](11$\bar{1}$) dislocations transform the crystal plane by plane, building up the twin. Blue regions represent the matrix, orange represents the growing twin.

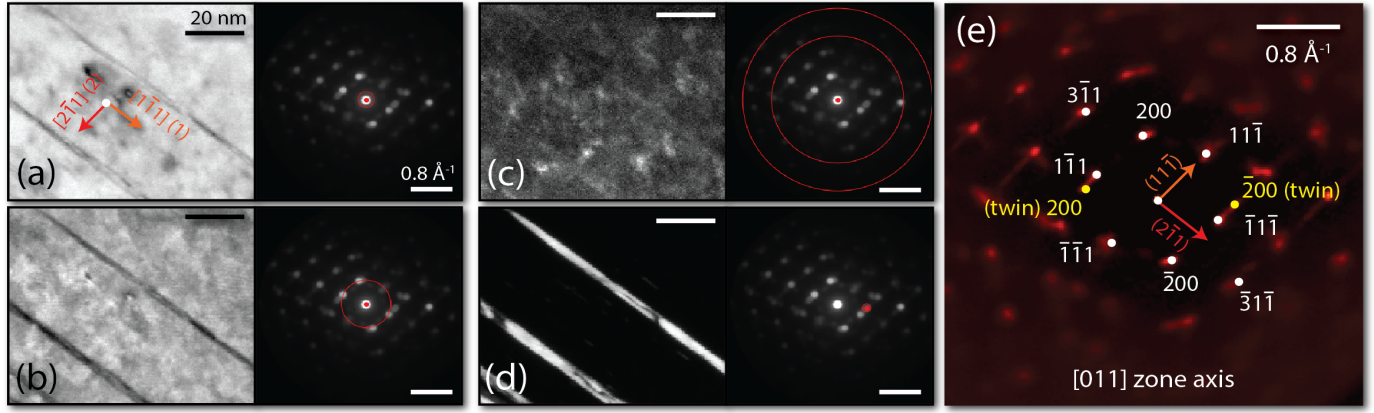

**Figure 2**: Virtual bright (a,b) and dark (c,d) field images of the steel nanotwin, as well as average Bragg vector maps (e) - sums over real space of the identified Bragg peak locations, weighted by intensity. The red circles in (a-d) diffraction patterns correspond to the limits of the virtual aperture, with the central points showing the centre of the virtual aperture..

A <110> zone was selected for this experiment, defined such that the [011] direction is out of the plane. EBSD was used to identify a suitable grain. An electron transparent sample was then prepared using the focussed ion beam (FIB) lift-out technique.

4D-STEM was performed using the probe and image spherical aberration-corrected *JEOL* ARM300CF TEM at *ePSIC*. The pencil beam was set-up by turning off the probe corrector hexapoles and using the condenser and transfer lens pairs to reach a small convergence semi-angle. Diffraction patterns were collected with a Merlin (MediPix) direct electron detector. An accelerating voltage of 200 kV and a camera length of 9 cm was used, with a 10 μm condenser lens aperture. Calibration diffraction data was gathered from evaporated gold on amorphous carbon sample and using the 10 μm aperture a 2.1 mrad convergence semi-angle and 0.0157 Å$^{-1}$ detector pixel size were measured. This gives a diffraction-limited spatial resolution of 0.73 nm. A 68.2-by-83.1 nm area of interest was scanned in 188-by-229 real space pixels, and with a 256-by-256 pixel diffraction pattern captured at each of these scan positions, with a 1 ms dwell time per pattern.

The in-plane elastic strain tensor was calculated from the electron diffraction pattern at each scan location. Bragg peak identification, dataset calibration including elliptical distortion correction and diffraction shift correction, and elastic strain calculation were performed with the open source py4DSTEM analysis package [8,22]. For locating Bragg peaks we use a correlation power of 1, corresponding to cross-correlation [13], and estimate the subpixel Bragg disc positions with local Fourier up-sampling by a factor of 16 [23].

The in-plane elastic stress was determined from the measured strains and Hooke's law. We make use of the axis system presented in Figure 1 (1, 2, 3) refer to the twinning system, with the (2, 3) plane at the interface (origin at the twin centre). In this scheme the [11$\bar{1}$] twin plane normal is aligned with the *2* direction, with *1* along [$\bar{2}$1$\bar{1}$]. To calculate the stresses, we first rotate a reference stiffness tensor, $\mathbf{C}^{ref}$ to our *1*, *2*, *3* axis system. We employ reference stiffness tensor components $C_{11}^{ref}$ = 197.5 GPa, $C_{12}^{ref}$ = 124.5 GPa, $C_{44}^{ref}$ = 122.0 GPa, with subscripts referring to tensor components rather than our axis systems, as measured for a similar austenite by Johansson *et al* [24,25] with respect to a <100> system. Having previously calculated the in-plane elastic strains resolved in the [1$\bar{2}$1] and [10$\bar{1}$] directions, we can infer the complete elastic stress state by assuming plane strain: we find the unknown strain components $\varepsilon_{33}$, $\varepsilon_{13}$, and $\varepsilon_{23}$ by assuming the stress components $\sigma_{33}$, $\sigma_{13}$, and $\sigma_{23}$ are zero. We then use the full strain vector and compliance tensor to calculate the full stress tensor, and the unknown, non-zero, $\sigma_{11}$, $\sigma_{22}$, and $\sigma_{12}$.

Presented in Figure 2 are 'virtual' bright and dark field (VBF, VDF respectively) images reconstructed from intensity collected from the highlighted digital apertures. VBF images (a) and (b) clearly distinguish the twin from the matrix. Given the axis system we have adopted, [011] out of plane, and the diffraction vectors as indexed in Figure 2, we infer these twins have habit plane (11$\bar{1}$).

A large amount of structure-dependent information is contained in the direct beam. Traditional bright field imaging (where an aperture isolates the direct beam) uses electron wave phase information to re-interfere an image [26]. In VBF we only have access to electron intensity in the diffraction plane, so it is likely that the contrast we observe between twin and matrix is derived from local strain, lattice rotation, or dynamical effects which will alter the ratio of diffracted to direct intensity. We present two VBF images Figure 2 (a,b) to show the presence of diffraction contrast in the direct beam as well as in the first order diffraction spots. The high angle VDF image in (c) is akin to conventional high-angle annular dark field (HAADF) imaging, which indicates propensity to scatter electrons to high angles, and suggests that there is no detectable variation in local chemistry between the twin and the matrix. The minimal contrast observed in (c) could be attributed to local ordering as a precursor to the formation of coherent V-rich carbides [27,28]. The twin is explicitly highlighted in (d) by reconstructing the spatial image from the $\bar{2}$00 reciprocal lattice point for the twinned region only, analogous to a traditional TEM dark field image. In (e) we present a Bragg vector map (after Savitzky *et al* [22]), representing a 2D probability histogram of peak locations in the untwinned region. We note that a small amount of density is observed at the twin reciprocal lattice points even for the untwinned class. This is possibly due to the geometry of the specimen, dynamical diffraction, and through-thickness sampling of both untwinned and twinned material near the interface.

The strain components $\varepsilon_{11}$, $\varepsilon_{22}$, $\varepsilon_{12}$ were calculated from the relative movements of the diffraction spots. This was performed independently for the twin and the matrix. A set of reference reciprocal lattice vectors were obtained by averaging the untwinned region's reciprocal lattice basis vectors. The twin basis vectors magnitudes were normalised to this unstrained length. As such, elastic strains are given in reference to this 'unstrained' state. The measurement could be considered as the elastic strain variation across the area of interest.

Maps of measured (11, 22, 12) and inferred (33) strain components across the area of interest are presented in Figure 3. Included is an integration along the 2-direction in a highlighted area, to obtain an average profile in the 1 direction. We observe that $\varepsilon_{22}$ and $\varepsilon_{12}$ remain fairly level in the matrix region between the twins, while $\varepsilon_{11}$ exhibits more deviation. The twinned regions exhibit an increase in $\varepsilon_{11}$ and $\varepsilon_{22}$, and a reduction in $\varepsilon_{12}$ relative to the matrix.

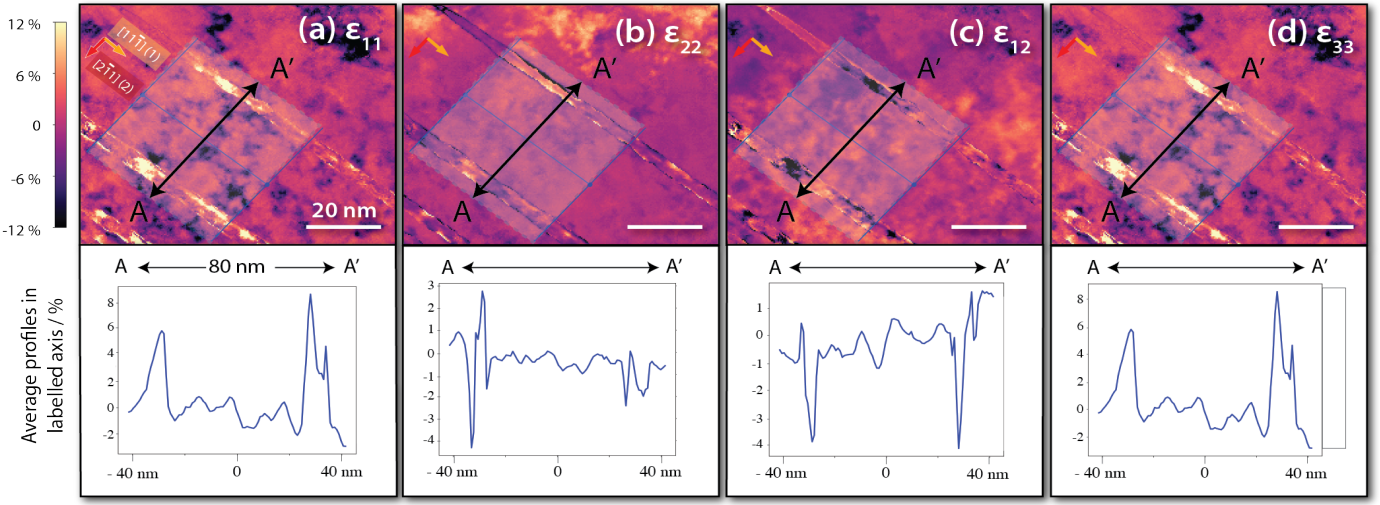

**Figure 3**: Elastic strain maps resolved in the 11 (a), 22 (b), 12 (c) and 33 (d) directions. An average 1-direction profile, perpendicular to the twin's length, was calculated by integrating all points in the 2-direction in the highlighted region.

Maps of the stress tensor are presented in Figure 4. The same line profile integration as for the strain maps is performed. We observe that $\sigma_{11}$ and $\sigma_{22}$ exhibit large positive stress rises across the interface. The component $\sigma_{12}$ is much smaller in magnitude, but generally observes a negative 'sense' shear stress profile.

For comparison to an analytical micromechanical model, we follow the formulation of Müllner et al [19–21]. This considers the elastic stress field around the twin to follow that of a disclination dipole. Given a distance $x$ along the 1-direction, and $y$ along the 2-direction, in this scheme:

$$\sigma_{11} = D\,\omega \left\{ \frac{1}{2} \log\left[\frac{x^2 + (y-a)^2}{x^2 + (y+a)^2}\right] + \frac{x^2}{x^2 + (y-a)^2} - \frac{x^2}{x^2 + (y+a)^2} \right\}$$

$$\sigma_{22} = D\,\omega \left\{ \frac{1}{2} \log\left[\frac{x^2 + (y-a)^2}{x^2 + (y+a)^2}\right] + \frac{x^2}{x^2 + (a)^2} - \frac{x^2}{x^2 + (y+a)^2} \right\}$$

$$\sigma_{12} = D\,\omega\, x \left\{ \frac{y+a}{x^2 + (y+a)^2} - \frac{y-a}{x^2 + (y-a)^2} \right\}$$

Using a natural logarithm, where $a$ is the twin half-thickness, with pre-factor $D$ given:

$$D = \frac{C_{44}^{ref}}{2\pi\,(1-\nu)}$$

Using the shear modulus, $\nu$ Poisson's ratio, and parameter $\omega$ the 'power' of the disclination:

$\omega = 2\,\tan^{-1}\frac{b}{2h} = 38.94°$ for FCC materials [20]

With $b$ as the magnitude of the Shockley partial Burgers vector and $h$ the separation between twinning planes. The co-ordinates $x, y, z$ are taken in reference to the centre of the disclination dipole from which the stress field is derived [20]. Models of $\sigma_{11}$, $\sigma_{22}$, and $\sigma_{12}$ along the **A-A'** profile are included in Figure 4. We set $y, z = 0$ at the centre of the twins, and vary $y$ along the profile **A-A'**. We find that an initial $x$ value of ~ -5 nm leads to stress profiles with similar form to those we observe. This corresponds to our sample representing an average of 5 nm along the $x$ direction away from the disclination dipole, which is the basis of the considered model. We use a Poisson's ratio of 0.31, and consider a twin thickness of

10 px (3.63 nm) [25]. Note that in order to qualitatively compare discretely sampled positions in the analytical model to our experiments, where we expect some degree of beam overlap, we apply a Chebyshev windowing function to the integration. This accelerates the function's descent towards zero in the limit, which was necessary for superposition of the two twin stress fields in our small area of interest.

Our strain measurements are significantly larger than previously seen in most 4D-STEM experiments, for example the ~4% range observed by Pekin et al [13]. Stresses of the magnitude we have measured are rarely observed under standard loading, but under conditions of severe plastic deformation in a drawn wire these levels are reached macroscopically [29].

Our sample has seen significant plastic strain. The tolerance for high defect density is precisely what makes TWIP steels the strongest and most readily work-hardening engineering materials available. The forms and sense of the stress profiles we have measured are similar to those predicted by the model. The (uniaxial) theoretical strength of a material can be approximated as Young's Modulus / 10 in the absence of accurate potentials [30]. This is around 20 GPa in our case. Our measurements lie below this threshold, but are close. The stresses predicted by the model appear to exceed our approximate threshold, but follow the same trend as the experimental results.

Koyama et al [18] have used electron channelling contrast imaging to investigate hydrogen embrittlement initiation at deformation twin boundaries. They observe that transgranular cracks always propagate along twin boundaries; we have measured $\sigma_{11}$ to be locally very high in a similar material, in accordance with their observation. Furthermore, here we provide evidence of significant strain at twin boundaries, which will attract the hydrogen and embrittle the steel, which Koyama et al set out as a softening mechanism. As they suggest, this is despite the coherence of the Σ3 boundary, and explains observed hydrogen trapping at such features in similar materials [31].

Finally, we suggest that the significant stress intensity observed parallel and perpendicular to their boundaries controls their thickening. Twin thickness is determined by the tendency to minimise elastic strain energy. Clausen et al [32] have described the twin internal back-stress generated by matrix constraint of the transformed twin in Mg: the plastic shear provided by the twinning transformation is spread over the incorporating grain, resulting in an equal and opposite elastic strain being contained within the twin. This elastic back-strain (leading to internal back-stress) is what we have observed. The twin thickness is controlled by accommodation of the transformation strain, as there is an energy penalty to this back-stress. The stress intensity surrounding our nanotwins thus prevents their thickening. This allows for a large number density of fine twins, enhanced grain sub-division, and a greater work hardening rate. As deformation progresses, plastic strain accumulation increases, resulting in a greater elastic back-stress. Equivalently, in larger grains where the twinning strain can be more widely distributed, the back stress is lower and twins are able to grow thicker, as suggested by Rahman *et al* [1,17].

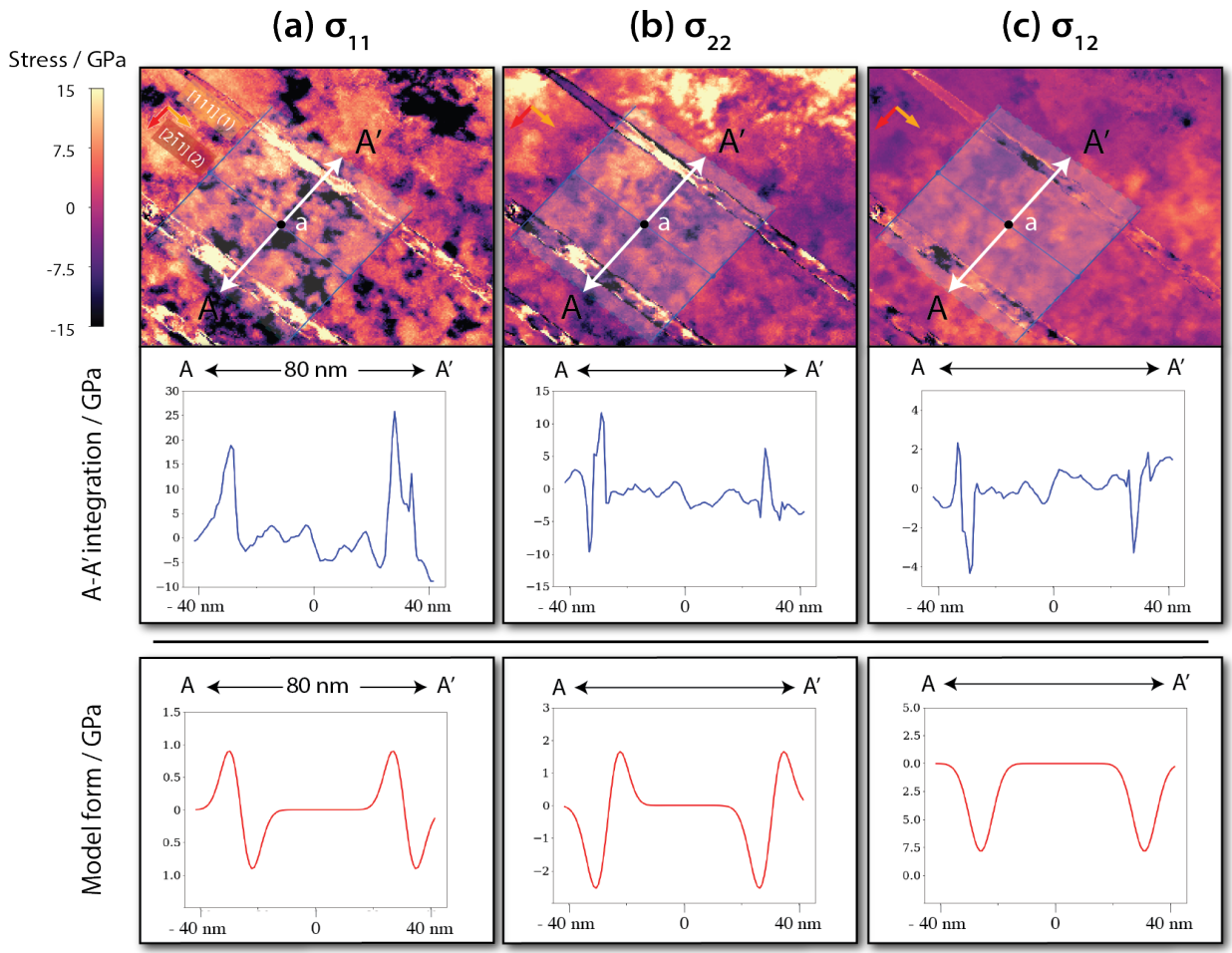

**Figure 4**: Elastic stress maps of the *11*, *22*, and *12* components. As in Figure 3, an integrated 1-direction profile (over the shaded region), perpendicular to the twin's length, was calculated by integrating all points in the 2-direction in the highlighted region.

In conclusion,

- Measurement of the elastic strain and stress state around deformation nanotwins in a TWIP steel reveals a significant polarisation, with stresses close to, but less than, the approximate theoretical strength of the material.

- The profiles and sense of the stresses follow those predicted by the analytical model of Müllner *et al* [19–21].

- The magnitude of the stress state surrounding the twin could explain the observations and provide evidence for hydrogen embrittlement mechanisms associated with twins set out by Koyama *et al* [18]. The observed stress field also likely plays a critical role in controlling twin thickness. This determines the rate of grain subdivision and the alloy work hardening rate [32].

**Authorship**: TPM co-acquired the data and drafted the initial manuscript. AKA co-acquired the data and FIB prepared the foil. TPM and BHS performed the analysis, using the py4DSTEM Python package developed by BHS and CO. TK developed the TWIP steel and plastically strained the sample. MD aligned, calibrated and supervised operation of the ePSIC TEM. CO provided critical microscopy and 4D-STEM insight. DD conceived the project and supervised the work. All the authors edited the manuscript.

**Acknowledgement**: TPM and DD would like to acknowledge support from the Rolls-Royce plc - EPSRC Strategic Partnership in Structural Metallic Systems for Gas Turbines (EP/M005607/1), and the Centre for Doctoral Training in Advanced Characterisation of Materials (EP/L015277/1) at Imperial College London. AA acknowledges EPSRC grant IAA EP/R511547/1. TK is grateful for support from A*STAR We thank Diamond Light Source for access and support in use of the electron Physical Science Imaging Centre (Instrument E02 and proposal number EM18770) that contributed to the results presented here. BHS and CO acknowledge funding from the Toyota Research Institute, and that work at the Molecular Foundry was supported by the Office of Science, Office of Basic Energy Sciences, of the U.S. Department of Energy under Contract No. DE-AC02-05CH11231. We are grateful to Ben Britton (Imperial) for insight into electron diffraction and micromechanics, and Ben Poole (Imperial) for guidance with basis rotation. We also thank Vassili Vorontsov (Strathclyde) and Alexander Knowles (Birmingham) for helpful conversations when planning the project.